\documentclass[aps,article,multicol,epsfig,showpacs]{revtex4}
\usepackage{amsfonts}
\usepackage{amsmath}
\usepackage{amssymb}
\usepackage{graphicx}%
\begin{document}

\preprint{HEP/123-qed}
\title{Nilpotent polynomials approach to four-qubit entanglement}
\author{Aikaterini Mandilara }
\affiliation{Laboratoire Aim\'{e} Cotton, B\^{a}t. 505, CNRS II, Campus d'Orsay, ORSAY\
CEDEX F-91405, FRANCE}
\author{Lorenza Viola }
\affiliation{Department of Physics and Astronomy,
Dartmouth College, 6127 Wilder Laboratory, Hanover, NH 03755, USA}


\begin{abstract}
We apply the general formalism of nilpotent polynomials [Mandilara
{\em et al}, {\em Phys. Rev. A} {\bf 74}, 022331 (2006)] to the
problem of pure-state multipartite entanglement classification in four
qubits.  In addition to establishing contact with existing results, we
explicitly show how the nilpotent formalism naturally suggests
constructions of entanglement measures invariant under the required
unitary or invertible class of local operations.  A candidate measure
of fourpartite entanglement is also suggested, and its behavior
numerically tested on random pure states.\end{abstract}

\startpage{1}

\maketitle

\section{Introduction}

Characterizing and quantifying multipartite entanglement is a problem
whose complexity rapidly increases with the number of particles, and a
major challenge within current quantum information science. In spite
of intensive effort, a complete understanding of entanglement
properties remains limited to date to few-body small-dimensional
composite quantum systems: in particular, such understanding has been
achieved for pure states of three two-level systems (qubits)
\cite{Dur,Gingrich}, mixed-state entanglement having also been
investigated for this system in \cite{Acin}.  Thus, the analysis of
pure-state entanglement in an ensemble of four qubits is a critical
test for any entanglement theory, as it is provides the first highly
non-trivial case whose complexity remains tractable.  Different
approaches have been attempted so far for unraveling the
classification of multipartite entanglement, ranging from so-called
hyper-determinants \cite{Akimasa}, to normal forms \cite{Verstraete},
and invariants \cite{Grassl,MLV,Levay2} and covariants \cite{Briand}
of the relevant group of local transformations.  Even if such methods
offer equivalent answers for ensembles of two or three qubits, a
complete description of four-qubit entanglement has only been obtained
by Verstraete \textit{et al }\cite{Verstraete2} based on the method of
normal forms, which simplifies considerably in this case thanks to the
fact that the group $SO(4,\mathbb{C})$ is isomorphic to $
SL(2,\mathbb{C})\times SL(2,\mathbb{C})$.  The resulting
classification has been partially independently verified in
\cite{Akimasa}.  Closely related with the problem of classification
is, in turn, the problem of quantifying entanglement through
appropriate measures, as the identification of proper classes should
provide the physical boundaries for prospective measures. In addition,
the invariants which are often utilized to discriminate among
different entanglement classes satisfy themselves the minimum set of
requirements that measures are expected to fulfill
\cite{Verstraete2,Vidal2}.

In this work, we tackle the problem of pure-state four-qubit
entanglement via a recently introduced approach based on {\em
nilpotent polynomials} \cite{nilpotent}.  In addition to providing a
simple entanglement criterion for any bipartition of a multipartite
ensemble, the nilpotent method has the advantage of offering, in
principle, a physically motivated procedure for entanglement
classification, based on the idea of reducing the nilpotent
polynomials to suitable {\em canonic} forms, which are {\em extensive}
with respect to the number of subsystems and {\em invariant} under the
desired groups of transformations.  Such a reduction procedure is
considerably facilitated if the dynamical equations of the polynomials
are derived and employed. The coefficients of the resulting invariant
forms have the same values as polynomial invariants, and may then be
used for constructing measures of entanglement.

The content of the paper is organized as follows.  After recalling in
Sec.~\ref{secI} the basic ingredients of the general nilpotent
formalism, we specialize it in Sec.~\ref{secII} to the four-qubit
setting, and derive both general and special entanglement classes for
this ensemble. Note that we obtain more entanglement classes than in
\cite{Verstraete2}, as a consequence of the fact that we consider at
each stage of our reduction procedure transformations that preserve
the canonic form of the nilpotent polynomials. In Sec.~\ref{secIII},
the problem of entanglement quantification is discussed in terms of
the invariant coefficients of the nilpotent polynomials. Measures for
comparing entanglement within classes are proposed, as well as a
measure of genuine fourpartite entanglement. Sec.~\ref{secIV}
concludes with a summary of the results, and a discussion of the main
advantages and limitations of our approach.

\section{Nilpotent polynomials for entanglement description}
\label{secI}

Consider a pure state $|\Psi\rangle$ describing an ensemble of $n$
qubits.  With respect to the computational basis in ${\cal H}\simeq
(\mathbb{C}^{2})^{\otimes n}$, $|\Psi\rangle$ may be expressed in the
form%
\begin{eqnarray}
\hspace*{-1cm}\left\vert \Psi \right\rangle &=&{\sum }_{\{k_{i}\}=0,1}\psi
_{k_{n}k_{n-1}\cdots k_{1}}\left\vert k_{n}k_{n-1}\cdots k_{1}\right\rangle 
\nonumber \\
\hspace*{-1cm} &=&\psi _{00\cdots 0}\left\vert 00\cdots 0\right\rangle +\psi
_{10\cdots 0}\left\vert 10\cdots 0\right\rangle +\ldots +\psi _{11\cdots
1}\left\vert 11\cdots 1\right\rangle ,  \label{EQ14}
\end{eqnarray}%
where $\psi_{k_{n}k_{n-1}\cdots k_{1}}\in \mathbb{C}$.  By
introducing pseudospin creation operators $\sigma_{i}^{+}$, the above
expression may be rewritten as
\begin{eqnarray*}
\hspace*{-1.5cm}\left\vert \Psi \right\rangle &=&\psi _{00\cdots 0}\left\vert
00\cdots 0\right\rangle +\psi _{10\cdots 0}\sigma _{n}^{+}\left\vert 0\cdots
0\right\rangle +\ldots +\psi _{11\cdots 1}\sigma _{n}^{+}\sigma
_{n-1}^{+}\cdots \sigma _{1}^{+}\left\vert 00\cdots 0\right\rangle \\
\hspace*{-1.5cm} &=&\bigg(\sum_{\{k_{i}\}=0,1}\psi _{k_{n}k_{n-1}\cdots
k_{1}}\prod_{i=1}^{n}\sigma _{i}^{+}\bigg)\left\vert 00\cdots 0
\right\rangle \:, 
\end{eqnarray*}%
that is, a polynomial in the nilpotent operator $\left\{ \sigma
_{i}^{+}\right\}$ (recall that $(\sigma_{i}^+)^{2}=0$ due to Pauli
algebra), acting on the vacuum (or reference) state $\left\vert
\mathrm{O}\right\rangle =\left\vert 00\cdots 0\right\rangle $. By
setting the population of the latter to be maximal (equal to one), we
construct, equivalently, the nilpotent polynomial $F$,
\begin{equation*}
\hspace*{-1.5cm}F(\{\sigma _{i}^{+}\})=\sum_{\left\{ k_{i}\right\} ={0,1}%
}\alpha _{k_{n}k_{n-1}\ldots k_{1}}\prod_{i=1}^{n}\left( \sigma
_{i}^{+}\right) ^{k_{i}}=\sum_{\{k_{i}\}={0,1}}\frac{\psi
_{k_{n}k_{n-1}\ldots k_{1}}}{\psi _{00\ldots 0}}\prod_{i=1}^{n}\left( \sigma
_{i}^{+}\right) ^{k_{i}}\ .\ 
\end{equation*}%
Furthermore, by taking the logarithm of $F$, and by Taylor-expanding
around the unit value of the vacuum state population, we obtain the
\textit{nilpotential }$f$,
\begin{equation*}
\hspace*{-1cm}f(\{\sigma _{i}^{+}\})=\ln \left[ F(\{\sigma _{i}^{+}\})\right]
=\sum_{\{k_{i}\}={0,1}}\beta _{k_{n}k_{n-1}\ldots
k_{1}}\prod_{i=1}^{n}\left( \sigma _{i}^{+}\right) ^{k_{i}}\ .
\end{equation*}%
The nilpotential makes it possible to readily check whether two
subsets $A$ and $B$ of qubits are entangled or not. The following
criterion holds \cite{nilpotent}:

\vspace{1mm}

\textbf{The entanglement criterion:} \textit{The subsets }$A$\textit{\
and }$B$ \textit{\ of a binary partition of an assembly of
}$n$\textit{\ qubits are unentangled iff}
\begin{equation*}
\frac{\partial ^{2}f(\{x_{i}\})}{\partial x_{k}\partial x_{m}}%
=0\,,\;\;\;\forall k\in A,\:\forall m\in B\:.
\end{equation*}%
Thus, $A$ and$~B$ are disentangled iff $ f_{A\cup B}=f_{A}(\left\{
x_{\in A}\right\} )+f_{B}(\left\{ x_{\in B}\right\} )$.  

\vspace{1mm}

In spite of the fact that the nilpotential $f$ gives the possibility
of applying the above entanglement criterion, $f$ may not yet be
regarded as a satisfactory description of entanglement present in the
overall composite system, as the latter should naturally be invariant
under operations which act {\em locally} on individual subsystems only
(see \cite{ge} for a generalization of entanglement {\em beyond} the
distinguishable subsystem framework we focus on here).  The local
transformations on each qubit may either be considered to be
restricted to unitary transformations in $SU(2)$ -- in which case, we
talk about $su$-entanglement -- or they may be more generally allowed
to be any invertible transformation in $SL(2,C)$ -- in which case, we
talk about $sl$-entanglement.  Physically, the latter correspond to
the family of {\em stochastic local operations assisted by classical
communication operations} ({SLOCC}) \cite{Bennett,Verstraete}.  Under
the action of local transformations (unitary or merely invertible),
the state vector undergoes changes but still remains within a subset
$\mathcal{O}$, which coincides with a $su$-orbit (or, respectively,
$sl$-orbit) within the overall Hilbert space $\mathcal{H}$.  Thus, the
nilpotential $f$ should retain the same form for all states belonging
to a given orbit, and a {\em canonic} form of the resulting
nilpotential may accordingly be taken as an ``orbit marker''. Canonic
forms may be used as an alternative to the method of invariants
\cite{Linden} for identifying different orbits, thereby entanglement
classes.  The number of independent (real) parameters in a given
canonic form should equal the number of independent invariants
identifying the orbit, or else equal the dimension of the coset
$\mathcal{H}/\mathcal{O}$.

According to the general arguments given in \cite{Carteret2,
nilpotent}, the {\em $su$-canonic nilpotential} is defined as the
nilpotential of the state in the orbit with the maximum reference
state population. Under this condition, the orbit-marker is the
canonic nilpotential, which we also term the \textit{su-tanglemeter},
$f_{c}$,
\begin{equation}
f_{c}(\{\sigma _{i}^{+}\})=\beta _{ij}\sigma _{i}^{+}\sigma _{j}^{+}+\ldots
,\;\;\;  \label{EQ13a}
\end{equation}%
where the $n$ linear terms are absent and the number of parameters
involved equals the dimension of the coset, $D_{su}=2^{n+1}-3n-2$, $n
\geq 3$.

In order to construct the {\em $sl$-canonic nilpotential}, or
$sl$-tanglemeter, we begin with the tanglemeter $f_{c}$, and we
further reduce the number of parameters down to $D_{sl}\ =\
2^{n+1}-6n-2$, $n \geq 4$.  To achieve this we impose the following
conditions: in addition to the requirement for $f_{c}$ that all $n$
terms linear in $\sigma_i^{+}$ be equal to zero, we require that
\emph{all $n$ terms of $(n-1)$-th order vanish as well}. Thus, the
$sl$-tanglemeter, $f_C$, takes the form
\begin{equation}
f_{C}(\{\sigma _{i}^{+}\})=\sum_{\quad \Sigma _{i}k_{i}\not \in
\{1,n-1\}}\beta _{k_{n}k_{n-1}\ldots k_{1}}\prod_{i=1}^{n}\left(
\sigma _{i}^{+}\right) ^{k_{i}}.  \label{EQ14a}
\end{equation}%
\bigskip 
Since $D_{sl}<D_{su}$, different $su$-orbits may become equivalent
under local $SL$-transformations.  For this reason, the classification
given by $SL$ is more general than the one given by $SU$, thus usually
the term \textquotedblleft entanglement classes\textquotedblright\ is
taken to refer to different $sl$-orbits.

Given an arbitrary pure state $|\Psi\rangle$, the task of determining
the tanglemeter by applying local operations is, in general, not
trivial. The difficulty is substantially reduced if one is able to
take advantage of explicit dynamical equation obeyed by the
nilpotential of the state, subject to appropriate consistency (or
``feedback'') conditions. For qubit systems, the dynamic equation
reads
\begin{equation}
\mathrm{i}\frac{\partial f}{\partial t}=\mathrm{e}^{-f}H\mathrm{e}^{f}\ ,
\label{EQ21af}
\end{equation}%
where the generators of the local operations%
\begin{equation}
H=\sum_{i}P_{i}^{-}(t)\sigma _{i}^{+}+P_{i}^{+}(t)\sigma
_{i}^{-}+P_{i}^{z}(t)\sigma _{i}^{z}\ ,  \label{EQ48}
\end{equation}%
should be formally substituted as
\begin{eqnarray}
\sigma _{i}^{+}f &=&\sigma _{i}^{+}f,  \nonumber \\
\sigma _{i}^{-}f &=&\frac{\partial f}{\partial \sigma _{i}^{+}}\,,
\label{EQ21aa} \\
\sigma _{i}^{z}f &=&-f+2\sigma _{i}^{+}\frac{\partial f}{\partial \sigma
_{i}^{+}}\ .  \nonumber
\end{eqnarray}

For the special case of local {\em unitary} operations, $\left(
P_{i}^{+}\right) ^{\ast }=\left( P_{i}^{-}\right)$ in
Eq.~(\ref{EQ48}), and the feedback conditions for obtaining $f_{c}$
are
\begin{equation}
P_{i}^{-}=\left( P_{i}^{+}\right) ^{\ast }=-\mathrm{i}\beta _{i}\,,
\label{EQ60}
\end{equation}%
where 
\begin{equation}
\beta _{i}=\left. \frac{\partial f}{\partial \sigma _{i}^{+}}\right\vert
_{\sigma \rightarrow 0}\,  \label{EQ58}
\end{equation}%
are the coefficients of the linear terms in the nilpotential at a
given time.

A similar procedure for reducing the nilpotential to the canonic form
$f_{C}$ may be carried out also for $SL$-transformations. We begin in
this case by reducing $f$ to the tanglemeter $f_{c}$, so that the
terms linear in $\sigma _{i}^{+}$ vanish.  Next, we apply $SL$
operations as in Eq.~(\ref{EQ48}), where however $P_{i}^{-}$ and
$P_{i}^{+}$ are no longer constrained to be complex conjugates, and
choose such operations in such a way that the terms in the
nilpotential involving the monomials of order one and of order $n-1$
in $\sigma_i^+$ decrease exponentially with time.  The two feedback
conditions to be imposed in this case are: ({i}) the condition
\begin{equation}
P_{j}^{-}=-\sum_{i=1}^{n}\left. P_{i}^{+}\frac{\partial ^{2}f}{\partial
\sigma _{i}^{+}\partial \sigma _{j}^{+}}\right\vert _{\sigma \rightarrow
0}=-\sum_{i=1}^{n}P_{i}^{+}\beta _{i,j}\:,  \label{EQ62a}
\end{equation}%
expressing $P_{i}^{-}$ via $P_{i}^{+}$, which ensures that the
nilpotential remains in the form of a tanglemeter at each stage; and
({ii}) the condition
\begin{equation}
\hspace*{-2cm}\left. \mathrm{i}\frac{\partial
^{n-1}f}{{\textstyle\prod_{i\neq j}}\partial \sigma
_{i}^{+}}\right\vert _{\sigma \rightarrow 0}=-P_{j}^{+}\frac{\partial
^{n}f}{{\textstyle\prod_{i}}\partial \sigma _{i}^{+}}
+\sum_{m=1}^{n}P_{k}^{+}\left. \frac{\partial ^{n-1}}{{\textstyle
\prod_{i\neq m}}\partial \sigma _{i}^{+}}\left[ \sigma _{i}^{+}\left(
\frac{\partial f}{\partial \sigma _{i}^{+}}\right) ^{2}\right]
\right\vert _{\sigma \rightarrow 0}, \label{EQ62b}
\end{equation}%
which ensures the exponential decrease of all $n$ coefficients in front of
the second-highest order terms.

Unfortunately, no immediate physical meaning seems to be attributable
in general to the requirements of vanishing of the $sl$-tanglemeter
coefficients of $\left( n-1\right) $-th order -- in contrast to the
case of $SU$ transformations, where vanishing of the first-order terms
reflects maximum ground state population.  Mathematically, however,
such a requirement is suggested by symmetry considerations: $n$
complex conditions are imposed on $n$ complex coefficients of the same
type. After having eliminated the monomials of orders $1$ and $(n-1)$,
it is possible to specify the scaling parameters $P_{i}^{z}$ so that
$n$ additional conditions are imposed on the tanglemeter coefficients.
For example, we can set to unity the coefficients in front of the
highest order term, and adjust $(n-1)$ coefficients in front of
certain monomials to be equal to $(n-1)$ coefficients of other
monomials.

The condition in Eq.~(\ref{EQ62b}) for $P_{j}^{+}$ is written
implicitly as a set of $n$ linear equations that can be solved for
{\em generic} states.  However, no solution exists for those
$P_{j}^{+}$ parameters corresponding to a zero determinant.  Such
singularities may correspond to {\em special} classes of entangled
states which require separate consideration -- as we are going to see
explicitly in the four-qubit example.

\section{ $sl$-tanglemeters for four qubits}
\label{secII}

A generic normalized pure state of four qubits may be described by
$2\cdot 2^{4}-2=30$ real parameters. The $su$-entanglement of this
state requires less parameters to be characterized, $D_{su}=30-3\cdot
4=18$ and, according to the discussion in Sec.~\ref{secI}, for four
qubits the $su$-tanglemeter defined in Eq.~(\ref{EQ13a}) reads
\begin{eqnarray}
\hspace{-1.8cm}f_{c} &=&\beta _{3}\sigma _{2}^{+}\sigma _{1}^{+}+\beta
_{5}\sigma _{3}^{+}\sigma _{1}^{+}+\beta _{9}\sigma _{4}^{+}\sigma
_{1}^{+}+\beta _{6}\sigma _{3}^{+}\sigma _{2}^{+}+\beta _{10}\sigma
_{4}^{+}\sigma _{2}^{+}+\beta _{12}\sigma _{4}^{+}\sigma _{3}^{+}
\label{EQ14.4} \\
\hspace{-1.8cm}&+&\beta _{7}\sigma _{3}^{+}\sigma _{2}^{+}\sigma
_{1}^{+}+\beta _{13}\sigma _{4}^{+}\sigma _{3}^{+}\sigma
_{1}^{+}+\beta _{11}\sigma _{4}^{+}\sigma _{2}^{+}\sigma
_{1}^{+}+\beta _{14}\sigma _{4}^{+}\sigma _{3}^{+}\sigma
_{2}^{+}+\beta _{15}\sigma _{4}^{+}\sigma _{3}^{+}\sigma
_{2}^{+}\sigma _{1}^{+}\,.  \nonumber
\end{eqnarray}%
In the above expression, we have used the local phase operations that
did not contribute to the elimination of the linear coefficients to
make the trilinear coefficients
$\beta_{7},\beta_{13},\beta_{11},\beta_{14}$ real numbers.  In
addition, a compact notation has been introduced by considering the
indexes of $\beta $ as a binary representation of decimal numbers,
e.g., $0011\mapsto 3$, \textit{etc}.

Allowing for more general local transformations on each qubit, such as
indirect measurements with stochastic outcomes, the number of the
parameters necessary to describe a state may be further reduced. The
$sl$-tanglemeter (\ref{EQ14a}) of a generic state of four qubits
contains $D_{sl}=30-6\cdot 4=6$ real parameters, and may be cast
in the following form:
\begin{eqnarray}
\hspace{-1.5cm}f_{C} &=&\beta _{3}\left( \sigma _{1}^{+}\sigma
_{2}^{+}+\sigma _{3}^{+}\sigma _{4}^{+}\right) +\beta _{5}\left( \sigma
_{1}^{+}\sigma _{3}^{+}+\sigma _{2}^{+}\sigma _{4}^{+}\right) +\beta
_{6}\left( \sigma _{1}^{+}\sigma _{4}^{+}+\sigma _{2}^{+}\sigma
_{3}^{+}\right)  \nonumber \\
\hspace{-1.5cm}&&+(1-\beta _{3}^{2}-\beta _{5}^{2}-\beta
_{6}^{2})\sigma _{1}^{+}\sigma _{2}^{+}\sigma _{3}^{+}\sigma _{4}^{+},
\label{EQ15}
\end{eqnarray}%
where the scaling factors (that is, the parameter in front of
$\sigma_{i}^{z}$ in Eq.~(\ref{EQ48})), have been chosen so that the
$f_{C}$ becomes equivalent to the expression $G_{abcd}$ in
\textit{Theorem 2} of \cite{Verstraete}.

We proceed to explicitly illustrate the procedure for evaluating the
$sl$-tanglemeter in Eq.~(\ref{EQ15}) by means of the dynamic equations
(\ref{EQ21af})-(\ref{EQ48}), starting from the $su$-tanglemeter given
in Eq.~(\ref{EQ14.4}). First, one may notice that in the system of
eleven first-order nonlinear differential equations for the
coefficients $\beta _{i}$, the coupling of the second-order terms
$\beta _{ij}\sigma _{i}^{+}\sigma _{j}^{+}$ to the fourth-order term $
\beta _{15}\sigma _{4}^{+}\sigma _{3}^{+}\sigma _{2}^{+}\sigma
_{1}^{+}$ occurs via the third-order terms $\beta _{7}\sigma
_{3}^{+}\sigma _{2}^{+}\sigma _{1}^{+}$, $\beta _{13}\sigma
_{4}^{+}\sigma _{3}^{+}\sigma _{1}^{+}$, $\beta _{11}\sigma
_{4}^{+}\sigma _{2}^{+}\sigma _{2}^{+}$, $ \beta _{14}\sigma
_{2}^{+}\sigma _{3}^{+}\sigma _{4}^{+}$. Thus, the time evolution of
all $\beta _{i}$ stops when these third-order coefficients $\beta
_{7}$, $\beta _{13}$, $\beta $, and $\beta _{14}$ vanish -- indicating
that for four qubits the $sl$-tanglemeter is a stationary solution for
the dynamic equations. If the coefficients $P_{i}^{-}$ satisfy the
requirement of Eq.~(\ref{EQ62a}), which ensures that the {\em
nilpotential always remains in the form of a valid $su$-tanglemeter
$f_{c}$ during such evolution}, what it is left is to adjust the time
dependence of the parameters $P_{1}^{+}$, $P_{2}^{+}$, $P_{3}^{+}$ and
$P_{4}^{+}$ so that they drive all four third-order coefficients to
zero.

From the differential equations of the third-order coefficients,%
\begin{eqnarray}
\mathrm{i}\dot{{\beta }}_{14} &=&\!-P_{1}^{+}\beta _{15}+2P_{2}^{+}\beta
_{6}\beta _{10}+2P_{3}^{+}\beta _{6}\beta _{12}+2P_{4}^{+}\beta _{10}\beta
_{12},  \nonumber \\
\mathrm{i}\dot{{\beta }}_{13} &=&2P_{1}^{+}\beta _{5}\beta
_{9}-P_{2}^{+}\beta _{15}+2P_{3}^{+}\beta _{5}\beta _{12}+2P_{4}^{+}\beta
_{9}\beta _{12},  \nonumber \\
\mathrm{i}\dot{{\beta }}_{11} &=&2P_{1}^{+}\beta _{3}\beta
_{9}+2P_{2}^{+}\beta _{3}\beta _{10}-P_{3}^{+}\beta _{15}+2P_{4}^{+}\beta
_{9}\beta _{10},  \nonumber \\
\mathrm{i}\dot{{\beta }}_{7} &=&2P_{1}^{+}\beta _{3}\beta
_{5}+2P_{2}^{+}\beta _{3}\beta _{6}+2P_{3}^{+}\beta _{5}\beta
_{6}-P_{4}^{+}\beta _{15}\,,  \label{EQ62d}
\end{eqnarray}%
we see that, in the general case, feedback conditions may be imposed
by a proper choice of the parameters $P_{i}^{+}$, in such a way that
these equations take the form
\begin{equation}
\dot{{\beta }}_{7}=-\beta _{7};\quad \dot{{\beta }}_{11}=-\beta _{11};\quad 
\dot{{\beta }}_{13}=-\beta _{13};\quad \dot{{\beta }}_{14}=-\beta _{14}\,.
\label{EQ62da}
\end{equation}%
The evolution implied by these equations brings, in turn, the
nilpotential to the following form:
\begin{eqnarray}
\hspace{-1mm}f&=&\beta _{3}\sigma _{2}^{+}\sigma _{1}^{+}+\beta _{5}\sigma
_{3}^{+}\sigma _{1}^{+}+\beta _{9}\sigma _{4}^{+}\sigma _{1}^{+}+\beta
_{6}\sigma _{3}^{+}\sigma _{2}^{+} \nonumber \\
\hspace{-1mm}& + &\beta _{10}\sigma _{4}^{+}\sigma
_{2}^{+}+\beta _{12}\sigma _{4}^{+}\sigma _{3}^{+}+\beta _{15}\sigma
_{4}^{+}\sigma _{3}^{+}\sigma _{2}^{+}\sigma _{1}^{+}\:.  \label{EQ14b}
\end{eqnarray}%
We can invoke the four scaling operators $e^{B_{i}\sigma _{i}^{z}}$,
and further reduce Eq.~(\ref{EQ14b}) to the $sl$-canonic form $f_{C}$
of Eq.~(\ref{EQ15}), unless one or more of the above $\beta $
coefficients vanish.  Such cases correspond to zero-measure manifolds
-- in other words, to special classes of entanglement. For example,
when $\beta _{3}=0$ in ~(\ref{EQ14b}), the tanglemeter may be cast, by
scaling, in the form
\begin{eqnarray}
f_{C}^{(2)} &=&\sigma _{3}^{+}\sigma _{4}^{+}+\beta _{5}(\sigma _{1}^{+}\sigma
_{3}^{+}+\sigma _{2}^{+}\sigma _{4}^{+})  \label{EQ16a} \\
&&+\beta _{6}(\sigma _{1}^{+}\sigma _{4}^{+}+\sigma _{2}^{+}\sigma
_{3}^{+})+(1-\beta _{5}^{2}-\beta _{6}^{2})\sigma _{1}^{+}\sigma
_{2}^{+}\sigma _{3}^{+}\sigma _{4}^{+}~,  \nonumber
\end{eqnarray}%
characterized by only two parameters. If $\beta _{3}=\beta _{10}=0$,
the $sl$-tanglemeter reads
\begin{equation}
\hspace*{-1cm}f_{C}^{(1)}=\sigma _{3}^{+}\sigma _{4}^{+}+\sigma
_{1}^{+}\sigma _{3}^{+}+\beta _{6}(\sigma _{1}^{+}\sigma
_{4}^{+}+\sigma _{2}^{+}\sigma _{3}^{+})+(1-\beta _{6}^{2})\sigma
_{1}^{+}\sigma _{2}^{+}\sigma _{3}^{+}\sigma _{4}^{+}~,
\label{EQ16b}
\end{equation}%
which only involves a single parameter.  Lastly, if $\beta_{3}=\beta
_{10}=\beta _{9}=0$,
\begin{equation}
f_{C}^{(0)}=\sigma _{3}^{+}\sigma _{4}^{+}+\sigma _{1}^{+}\sigma
_{3}^{+}+\sigma _{2}^{+}\sigma _{3}^{+}+\sigma _{1}^{+}\sigma
_{2}^{+}\sigma _{3}^{+}\sigma _{4}^{+}~.  \label{EQ16c}
\end{equation}%
Note that the tanglemeters of Eqs.~(\ref{EQ16a}), (\ref{EQ16b}), and
(\ref{EQ16c}) correspond to the special families $L_{abc_{2}}$,
$L_{a_{2}b_{2}}$ and $ L_{a_{2}0_{3\oplus 1}}$ of the classification
given in \textit{Theorem 2} of \cite{Verstraete}. However, it is
important to bear in mind that the latter classification applies to
{\em un}-normalized states, whereas our tanglemeter corresponds to
states of unit population in the reference state.

When the fourth-order coefficient $\beta _{15}=0$ and, additionally,
one or more of the quadratic coefficients are also zero, singular
classes of states without genuine fourpartite entanglement emerge: for
instance, the $sl$-tanglemeter of a four-qubit $W$ state,
\begin{equation*}
f_{C}=\sigma _{3}^{+}\sigma _{4}^{+}+\sigma _{1}^{+}\sigma _{3}^{+}+\sigma
_{2}^{+}\sigma _{3}^{+},
\end{equation*}%
belongs to one of such classes, and separable states with tanglemeters
of the type
\begin{equation*}
f_{C}=\sigma _{3}^{+}\sigma _{4}^{+}+\sigma _{2}^{+}\sigma _{3}^{+}+\sigma
_{2}^{+}\sigma _{4}^{+}~,
\end{equation*}%
and similar, belong to other.

On the other hand, reducing $\ f_{c}$ to the canonic $sl$-form $f_{C}$
cannot be achieved when the determinant
\begin{equation}
{\cal D}_4=\left\vert
\begin{array}{cccc}
-\beta _{15} & 2\beta _{6}\beta _{10} & 2\beta _{6}\beta _{12} & 2\beta
_{10}\beta _{12} \\ 
2\beta _{5}\beta _{9} & -\beta _{15} & 2\beta _{5}\beta _{12} & 2\beta
_{9}\beta _{12} \\ 
2\beta _{3}\beta _{9} & 2\beta _{3}\beta _{10} & -\beta _{15} & 2\beta
_{9}\beta _{10} \\ 
2\beta _{3}\beta _{5} & 2\beta _{3}\beta _{6} & 2\beta _{5}\beta _{6} & 
-\beta _{15}%
\end{array}%
\right\vert  \label{EQ62e}
\end{equation}%
of the system of differential equations (\ref{EQ62d}) vanishes --
which makes it impossible to impose any required feedback
conditions. In such a situation, we loose the functional independence
of the right hand sides of (\ref{EQ62d}), which ensures complete
controllability of the dynamics of $\beta _{7}$, $\beta _{13}$, $\beta
_{11}$, and $\beta _{14}$ in the generic case. In turn, this means
that some linear combinations of these coefficients, determined by the
system's eigenvectors, cannot be set to zero by any choice of
$P_{i}^{+}$, and a tanglemeter $f_{C}$ of a special form should be
defined in such instances. In \cite{nilpotent}, four special families
of tanglemeters are derived,%
\begin{eqnarray}
f_{C}^{(s1)} &=&\beta _{3}\left( \sigma _{2}^{+}\sigma _{1}^{+}+\sigma
_{4}^{+}\sigma _{3}^{+}\right) +\beta _{5}\left( \sigma _{3}^{+}\sigma
_{1}^{+}+\sigma _{4}^{+}\sigma _{2}^{+}\right)  \nonumber \\
&&+\beta _{6}\left( \sigma _{4}^{+}\sigma _{1}^{+}+\sigma _{3}^{+}\sigma
_{2}^{+}\right)  \nonumber \\
&&+\sigma _{3}^{+}\sigma _{2}^{+}\sigma _{1}^{+}-\sigma _{4}^{+}\sigma
_{2}^{+}\sigma _{1}^{+}+\sigma _{4}^{+}\sigma _{3}^{+}\sigma _{1}^{+}-\sigma
_{4}^{+}\sigma _{3}^{+}\sigma _{2}^{+}  \nonumber \\
&&+2\left( \beta _{5}\beta _{6}-\beta _{3}\beta _{6}+\beta _{3}\beta
_{5}\right) \sigma _{4}^{+}\sigma _{3}^{+}\sigma _{2}^{+}\sigma _{1}^{+}\,,
\label{EQ14.4a}
\end{eqnarray}
\begin{eqnarray}
f_{C}^{(s2)} &=&\sigma _{2}^{+}\sigma _{1}^{+}+\sigma _{3}^{+}\sigma
_{1}^{+}+\sigma _{4}^{+}\sigma _{2}^{+}+\sigma _{4}^{+}\sigma _{3}^{+} 
\nonumber \\
&&+\beta _{6}\left( \sigma _{4}^{+}\sigma _{1}^{+}+\sigma _{3}^{+}\sigma
_{2}^{+}\right) +\beta _{7}\left( \sigma _{3}^{+}\sigma _{2}^{+}\sigma
_{1}^{+}-\sigma _{4}^{+}\sigma _{3}^{+}\sigma _{2}^{+}\right)  \nonumber \\
&&+\beta _{11}\left( \sigma _{4}^{+}\sigma _{2}^{+}\sigma _{1}^{+}-\sigma
_{4}^{+}\sigma _{3}^{+}\sigma _{1}^{+}\right) +2\sigma _{4}^{+}\sigma
_{3}^{+}\sigma _{2}^{+}\sigma _{1}^{+}\,.  \label{EQ14.4b}
\end{eqnarray}%
\begin{eqnarray}
f_{C}^{(s3)} &=&\sigma _{2}^{+}\sigma _{1}^{+}+\sigma _{3}^{+}\sigma
_{1}^{+}+\sigma _{4}^{+}\sigma _{2}^{+}+\sigma _{4}^{+}\sigma
_{3}^{+}+\sigma _{4}^{+}\sigma _{1}^{+}+\sigma _{3}^{+}\sigma _{2}^{+} 
\nonumber \\
&&+\beta _{14}\left( \sigma _{3}^{+}\sigma _{2}^{+}\sigma _{1}^{+}-\sigma
_{4}^{+}\sigma _{2}^{+}\sigma _{1}^{+}+\sigma _{4}^{+}\sigma _{3}^{+}\sigma
_{1}^{+}-\sigma _{4}^{+}\sigma _{3}^{+}\sigma _{2}^{+}\right)  \nonumber \\
&&+\beta _{13}\left( \sigma _{3}^{+}\sigma _{2}^{+}\sigma _{1}^{+}+\sigma
_{4}^{+}\sigma _{2}^{+}\sigma _{1}^{+}-\sigma _{4}^{+}\sigma _{3}^{+}\sigma
_{1}^{+}-\sigma _{4}^{+}\sigma _{3}^{+}\sigma _{2}^{+}\right)  \nonumber \\
&&+\beta _{11}\left( \sigma _{3}^{+}\sigma _{2}^{+}\sigma _{1}^{+}-\sigma
_{4}^{+}\sigma _{2}^{+}\sigma _{1}^{+}-\sigma _{4}^{+}\sigma _{3}^{+}\sigma
_{1}^{+}+\sigma _{4}^{+}\sigma _{3}^{+}\sigma _{2}^{+}\right)  \nonumber \\
&&+2\sigma _{4}^{+}\sigma _{3}^{+}\sigma _{2}^{+}\sigma _{1}^{+}\,.
\label{EQ14.44c}
\end{eqnarray}%
\begin{eqnarray}
f_{C}^{(s4)} &=&\sigma _{3}^{+}\sigma _{2}^{+}\sigma _{1}^{+}+\sigma
_{4}^{+}\sigma _{3}^{+}\sigma _{1}^{+}+\sigma _{4}^{+}\sigma
_{2}^{+}\sigma _{1}^{+}+\sigma _{4}^{+}\sigma _{3}^{+}\sigma _{2}^{+}
\nonumber \\ &&+\beta _{3}\sigma _{2}^{+}\sigma _{1}^{+}+\beta
_{5}\sigma _{3}^{+}\sigma _{1}^{+}+\beta _{6}\sigma _{3}^{+}\sigma
_{2}^{+}, \label{EQ14.44d}
\end{eqnarray}%
corresponding, respectively, to one, two, three, or four of the
eigenvalues $\gamma_j$ of ${\cal D}_4$ vanishing, where explicitly 
\begin{eqnarray}
\gamma_{1} &=&\beta _{15}-2\sqrt{\beta _{5}\beta _{6}\beta _{9}\beta _{10}}%
+2\sqrt{\beta _{3}\beta _{6}\beta _{9}\beta _{12}}-2\sqrt{\beta _{3}\beta
_{5}\beta _{10}\beta _{12}}\,,  \nonumber \\
\gamma_{2} &=&\beta _{15}+2\sqrt{\beta _{5}\beta _{6}\beta _{9}\beta _{10}}%
-2\sqrt{\beta _{3}\beta _{6}\beta _{9}\beta _{12}}-2\sqrt{\beta _{3}\beta
_{5}\beta _{10}\beta _{12}}\,,  \nonumber \\
\gamma_{3} &=&\beta _{15}-2\sqrt{\beta _{5}\beta _{6}\beta _{9}\beta _{10}}%
-2\sqrt{\beta _{3}\beta _{6}\beta _{9}\beta _{12}}+2\sqrt{\beta _{3}\beta
_{5}\beta _{10}\beta _{12}}\,,  \nonumber \\
\gamma_{4} &=&\beta _{15}+2\sqrt{\beta _{5}\beta _{6}\beta _{9}\beta _{10}}%
+2\sqrt{\beta _{3}\beta _{6}\beta _{9}\beta _{12}}+2\sqrt{\beta _{3}\beta
_{5}\beta _{10}\beta _{12}}\,. \label{EQ62.1}
\end{eqnarray}%
Observe that the number of parameters in such special tanglemeters is
$3$ complex numbers, the same as in the general case given in
Eq.~(\ref{EQ15}).

At present, it remains to be proved whether all four special
tanglemeters (\ref{EQ14.4a})-(\ref{EQ14.44d}) correspond to distinct
special entanglement classes, since they result from considering a
dynamic evolution based on a series of sequential {\em infinitesimal}
local operations which preserve the $su$-canonic form of the
nilpotential. Thus, a situation where some of the obtained
tanglemeters turn out to be equivalent under a {\em finite} local
$sl$-transformation, cannot be ruled out in principle by our current
approach. In \cite{Verstraete}, such special classes are not
explicitly identified, although the last three classes of
\textit{Theorem 2} -- that is, $L_{0_{7\oplus \bar{1}}}$ ,
$L_{0_{3\oplus \bar{1}}0_{3\oplus \bar{1}}}$, and $L_{0_{5\oplus
\bar{3}}}$, may be easily identified as special cases of
Eq.~(\ref{EQ14.44d}) when one or more terms vanish. The class
$L_{ab_{3}}$ in \cite{Verstraete} \ is not identified by our method.

We summarize in Table 1 the entanglement classes for pure states of
four qubits we have thus obtained.

\begin{table}[tbp]
\small
\label{tables} 
\begin{tabular}{cc}
\emph{General class} & 3 complex parameters \\ \hline
$G_{a}$ & $f=\beta _{3}(\sigma _{1}^{+}\sigma _{2}^{+}+\sigma _{3}^{+}\sigma
_{4}^{+})+\beta _{5}(\sigma _{1}^{+}\sigma _{3}^{+}+\sigma _{2}^{+}\sigma
_{4}^{+})$ \\ 
& $+\beta _{6}(\sigma _{1}^{+}\sigma _{4}^{+}+\sigma _{2}^{+}\sigma
_{3}^{+})+(1-\beta _{3}^{2}-\beta _{5}^{2}-\beta _{6}^{2})\sigma
_{1}^{+}\sigma _{2}^{+}\sigma _{3}^{+}\sigma _{4}^{+}$ \\ 
\emph{Singular $3D$ classes} & 3 complex parameters \\ \hline
$G_{b}$ & $f=\beta _{3}\left( \sigma _{1}^{+}\sigma _{2}^{+}+\sigma
_{3}^{+}\sigma _{4}^{+}\right) +\beta _{5}\left( \sigma _{1}^{+}\sigma
_{3}^{+}+\sigma _{2}^{+}\sigma _{4}^{+}\right) +\beta _{6}\left( \sigma
_{1}^{+}\sigma _{4}^{+}+\sigma _{2}^{+}\sigma _{3}^{+}\right) $ \\ 
& $+\sigma _{1}^{+}\sigma _{2}^{+}\sigma _{3}^{+}-\sigma _{1}^{+}\sigma
_{2}^{+}\sigma _{4}^{+}+\sigma _{1}^{+}\sigma _{3}^{+}\sigma _{4}^{+}-\sigma
_{2}^{+}\sigma _{3}^{+}\sigma _{4}^{+}$ \\ 
& $+2\left( \beta _{5}\beta _{6}-\beta _{3}\beta _{6}+\beta _{3}\beta
_{5}\right) \sigma _{1}^{+}\sigma _{2}^{+}\sigma _{3}^{+}\sigma _{4}^{+}$ \\ 
&  \\ 
$G_{c}$ & $f=\sigma _{1}^{+}\sigma _{2}^{+}+\sigma _{1}^{+}\sigma
_{3}^{+}+\sigma _{2}^{+}\sigma _{4}^{+}+\sigma _{3}^{+}\sigma _{4}^{+}+\beta
_{6}\left( \sigma _{1}^{+}\sigma _{4}^{+}+\sigma _{2}^{+}\sigma
_{3}^{+}\right) $ \\ 
& $+\beta _{7}\left( \sigma _{1}^{+}\sigma _{2}^{+}\sigma _{3}^{+}-\sigma
_{2}^{+}\sigma _{3}^{+}\sigma _{4}^{+}\right) +\beta _{11}\left( \sigma
_{1}^{+}\sigma _{2}^{+}\sigma _{4}^{+}-\sigma _{1}^{+}\sigma _{3}^{+}\sigma
_{4}^{+}\right) $ \\ 
& $+2\sigma _{1}^{+}\sigma _{2}^{+}\sigma _{3}^{+}\sigma _{4}^{+}$ \\ 
&  \\ 
$G_{d}$ & $f=\sigma _{1}^{+}\sigma _{2}^{+}+\sigma _{1}^{+}\sigma
_{3}^{+}+\sigma _{2}^{+}\sigma _{4}^{+}+\sigma _{3}^{+}\sigma
_{4}^{+}+\sigma _{1}^{+}\sigma _{4}^{+}+\sigma _{2}^{+}\sigma _{3}^{+}$ \\ 
& $+\beta _{14}\left( \sigma _{1}^{+}\sigma _{2}^{+}\sigma _{3}^{+}-\sigma
_{1}^{+}\sigma _{2}^{+}\sigma _{4}^{+}+\sigma _{1}^{+}\sigma _{3}^{+}\sigma
_{4}^{+}-\sigma _{2}^{+}\sigma _{3}^{+}\sigma _{4}^{+}\right) $ \\ 
& $+\beta _{13}\left( \sigma _{1}^{+}\sigma _{2}^{+}\sigma _{3}^{+}+\sigma
_{1}^{+}\sigma _{2}^{+}\sigma _{4}^{+}-\sigma _{1}^{+}\sigma _{3}^{+}\sigma
_{4}^{+}-\sigma _{2}^{+}\sigma _{3}^{+}\sigma _{4}^{+}\right) $ \\ 
& $+\beta _{11}\left( \sigma _{1}^{+}\sigma _{2}^{+}\sigma _{3}^{+}-\sigma
_{1}^{+}\sigma _{2}^{+}\sigma _{4}^{+}-\sigma _{1}^{+}\sigma _{3}^{+}\sigma
_{4}^{+}+\sigma _{2}^{+}\sigma _{3}^{+}\sigma _{4}^{+}\right) $ \\ 
& $+2\sigma _{1}^{+}\sigma _{2}^{+}\sigma _{3}^{+}\sigma _{4}^{+}$ \\ 
&  \\ 
$G_{e}$ & $f=\sigma _{1}^{+}\sigma _{2}^{+}\sigma _{3}^{+}+\sigma
_{1}^{+}\sigma _{2}^{+}\sigma _{4}^{+}+\sigma _{1}^{+}\sigma _{3}^{+}\sigma
_{4}^{+}+\sigma _{2}^{+}\sigma _{3}^{+}\sigma _{4}^{+}+$ \\ 
& $\beta _{3}\sigma _{1}^{+}\sigma _{2}^{+}+\beta _{6}\sigma _{2}^{+}\sigma
_{3}^{+}+\beta _{5}\sigma _{2}^{+}\sigma _{3}^{+}$ \\ 
&  \\ 
\emph{Singular $2D$ classes} & 2 complex parameters \\ \hline
&  \\ 
$LG2_{a}$ & $f=\sigma _{3}^{+}\sigma _{4}^{+}+\beta _{5}(\sigma
_{1}^{+}\sigma _{3}^{+}+\sigma _{2}^{+}\sigma _{4}^{+})+\beta _{6}(\sigma
_{1}^{+}\sigma _{4}^{+}+\sigma _{2}^{+}\sigma _{3}^{+})+(1-\beta
_{5}^{2}-\beta _{6}^{2})\sigma _{1}^{+}\sigma _{2}^{+}\sigma _{3}^{+}\sigma
_{4}^{+}$ \\ 
&  \\ 
$LG2_{b}$ & $f=\sigma _{1}^{+}\sigma _{2}^{+}+\sigma _{3}^{+}\sigma
_{4}^{+}+\beta _{5}(\sigma _{1}^{+}\sigma _{3}^{+}+\sigma _{2}^{+}\sigma
_{4}^{+})+\beta _{6}(\sigma _{1}^{+}\sigma _{4}^{+}+\sigma _{2}^{+}\sigma
_{3}^{+})$ \\ 
&  \\ 
$LG2_{c}$ & $f=\sigma _{1}^{+}\sigma _{3}^{+}\sigma _{4}^{+}+\sigma
_{1}^{+}\sigma _{2}^{+}\sigma _{4}^{+}+\sigma _{2}^{+}\sigma _{3}^{+}\sigma
_{4}^{+}+\sigma _{1}^{+}\sigma _{2}^{+}+\beta _{5}\sigma _{1}^{+}\sigma
_{3}^{+}+\beta _{6}\sigma _{2}^{+}\sigma _{3}^{+}$ \\ 
&  \\ 
\emph{Singular $1D$ classes} & 1 complex parameters \\ \hline
&  \\ 
$LG1_{a} $ & $f=\sigma _{1}^{+}\sigma _{2}^{+}+\sigma _{1}^{+}\sigma
_{3}^{+}+\beta _{6}(\sigma _{1}^{+}\sigma _{4}^{+}+\sigma _{2}^{+}\sigma
_{3}^{+})+(1-\beta _{6}^{2})\sigma _{1}^{+}\sigma _{2}^{+}\sigma
_{3}^{+}\sigma _{4}$ \\ 
&  \\ 
$LG1_{b}$ & $f=\sigma _{1}^{+}\sigma _{2}^{+}\sigma _{4}^{+}+\sigma
_{2}^{+}\sigma _{3}^{+}\sigma _{4}^{+}+\sigma _{1}^{+}\sigma _{2}^{+}+\sigma
_{1}^{+}\sigma _{3}^{+}+\beta _{6}\sigma _{2}^{+}\sigma _{3}^{+}$ \\ 
&  \\ 
\emph{Singular point classes} & no parameters \\ \hline
$S_{a}$ & $f=\sigma _{1}^{+}\sigma _{2}^{+}\sigma _{3}^{+}+\sigma
_{1}^{+}\sigma _{3}^{+}\sigma _{4}^{+}+\sigma _{2}^{+}\sigma _{4}^{+}$ \\ 
$S_{b}$ & $f=\sigma _{1}^{+}\sigma _{2}^{+}\sigma _{3}^{+}+\sigma
_{1}^{+}\sigma _{3}^{+}\sigma _{4}^{+}+\sigma _{1}^{+}\sigma _{2}^{+}\sigma
_{4}^{+}$ \\ 
$S_{c}$ & $f=\sigma _{1}^{+}\sigma _{2}^{+}\sigma _{3}^{+}+\sigma
_{1}^{+}\sigma _{3}^{+}\sigma _{4}^{+}$ \\ 
$S_{d}$ & $f=\sigma _{1}^{+}\sigma _{2}^{+}\sigma _{3}^{+}$ \\ 
$S_{e}$ & $f=\sigma _{3}^{+}\sigma _{4}^{+}+\sigma _{1}^{+}\sigma
_{3}^{+}+\sigma _{2}^{+}\sigma _{3}^{+}+\sigma _{1}^{+}\sigma _{2}^{+}\sigma
_{3}^{+}\sigma _{4}^{+}$ \\ 
$S_{f}$ & $f=\sigma _{1}^{+}\sigma _{2}^{+}+\sigma _{2}^{+}\sigma
_{3}^{+}+\sigma _{3}^{+}\sigma _{1}^{+}+\sigma _{1}^{+}\sigma _{2}^{+}\sigma
_{3}^{+}\sigma _{4}^{+}$ \\ 
$\dots $ & $\dots $ \\ 
& 
\end{tabular}%
\normalsize
\caption{Classification of four-qubit entanglement classes following
from $ SL(2,{\mathbb C})$ transformation properties of the canonic
form, see Sec. 3.}
\end{table}

\section{Entanglement measures for four qubits}
\label{secIII}

\subsection{Measures for $sl$- and $su$-entanglement}

From an information-theoretic standpoint, the construction of
well-defined entanglement measures typically relies on the concept of
{\em entanglement monotone}, that is, of a quantity that is required
to be invariant under local unitary transformations and non-increasing
on average under LOCC transformations \cite{Bennett}. For instance,
the most widely utilized measures for two and three qubits, the
\textit{concurrence}, $C$, and the \textit{residual entanglement} (or
3-tangle), $\tau$, \cite{Wootters} are entanglement monotones. For a
four-qubit system, we have seen in Sec. \ref{secII} that the
classification is much richer than in the case of three qubits. In the
context of such a classification, we would like to first revisit the
role of entanglement monotones, and then argue that another class of
measures may also be meaningful. In particular, we show how a measure
for fourpartite entanglement should be also more precisely defined by
imposing additional requirements beside the ones mentioned above.

A standard way to construct entanglement monotones is based on
exploiting polynomial (algebraic) invariants. Polynomial invariants
are polynomial functions of the state coefficients, and a linearly
independent finite set of them may be used to distinguish different
orbits in the same way the set of invariant tanglemeter's coefficients
does. For example, for a three-qubit system, five (as many as the
tanglemeter's parameters) independent invariants under local unitary
transformations exist \cite{Gingrich}, namely the three real numbers
\begin{eqnarray}
I_{1}& =\psi _{kij}\psi ^{\ast pij}\psi _{pmn}\psi ^{\ast kmn}\ ,
\nonumber \\ I_{2}& =\psi _{ikj}\psi ^{\ast ipj}\psi _{mpn}\psi ^{\ast
mkn\ }, \nonumber \\ I_{3}& =\psi _{ijk}\psi ^{\ast ijp}\psi
_{mnp}\psi ^{\ast mnk\ },
\label{EQ3}
\end{eqnarray}%
and the real and the imaginary part of a complex number, 
\begin{equation}
I_{4}+\mathrm{i}I_{5}=\psi _{ijk}\psi ^{ijp}\psi _{mnp}\psi ^{mnk}\,.
\label{EQ3a}
\end{equation}%
In the above equations, $\psi ^{ijk}=\epsilon ^{ii^{\prime }}\epsilon
^{jj^{\prime }}\epsilon ^{kk^{\prime }}\psi _{i^{\prime }j^{\prime
}k^{\prime }}$, with the convention that summation over repeated
indexes ranging over $\{0,1\}$ is left implicit, $\psi ^{\ast
}{}^{ijk}$ denotes the complex conjugate of $\psi _{ijk}$, and
$\epsilon ^{ii^{\prime }}$ is the antisymmetric tensor of rank
$2$. The $su$-invariant quantity $\tau=2|I_{4}+iI_{5}|$ is exactly the
3-tangle, which also remains invariant under the class of local
transformations $\otimes _{i=1}^{3}SL_{i}(2,\mathbb{C})$. It was
proved in \cite{Verstraete} that $sl$-invariants behave as
entanglement monotones for normalized pure states, since the (square)
vector length $\sum_i \psi _{i}\psi _{i}^{\ast}$ is non-increasing
under $sl$-transformations and thus may be employed as a measure of
$su$-entanglement within a given $sl$ orbit.  The main reason for
choosing such a vector length as a measure is based on the relation
between the determinant $D_{et}\leq 1$ of the physical transformation
corresponding to the chosen $sl$-transformation and the probability
$\left(\sum_i \psi _{i}\psi _{i}^{\ast}\right)^{-1}$ of the desired
outcome of the indirect measurement implementing this transformation:
The $\left( n-1\right)$-th power of the probability upper-bounds the
determinant, $D_{et}\leq \left( \sum_i \psi _{i}\psi _{i}^{\ast
}\right) ^{-n}$.

Note that, by definition, an entanglement monotone is an object able
to quantify $su$-entanglement by distinguishing different $su$-orbits
that belong to the same $sl$-orbit. However, in the case of four (or
more) qubits, there exists an infinite number of general $sl$-orbits
(see Sec. \ref{secII}). This suggests that measures able to compare
the {\em $sl$-entanglement between such general orbits} should be
considered in addition to the $su$-measures.  A reasonable suggestion
for $sl$-entanglement measures is provided by $sl$-invariants that are
also scaling invariants and, therefore, are independent of the
specific normalization of the state. One may construct $sl$-invariants
for a four-qubit ensemble in a way similar to how the invariant
$I_{4}+iI_{5}$ of Eq.~(\ref{EQ3a}) is constructed; that is, by taking
products of several factors $\sim \psi $ (but not factors $\sim \psi
^{\ast }$) and by considering contractions over $SU(2)$-indexes with
invariant antisymmetric tensors $\epsilon ^{ii^{\prime }}$. The
simplest combination one finds in this way,
\begin{equation}
I^{(2)}\ =\ \psi _{ijkl}\psi ^{ijkl}\,,  \label{J2}
\end{equation}%
is a $sl$-invariant of second order. There also exist three different
$sl$-invariants of fourth order,
\begin{eqnarray}
I_{12}^{(4)}& =I_{34}^{(4)}\ =\ \psi _{ijkl}\psi ^{ijmn}\psi _{opmn}\psi
^{opkl}\,,  \nonumber \\
I_{13}^{(4)}& =I_{24}^{(4)}\ =\ \psi _{ikjl}\psi ^{imjn}\psi _{ompn}\psi
^{okpl}\,,  \nonumber \\
I_{14}^{(4)}& =I_{23}^{(4)}\ =\ \psi _{iklj}\psi ^{imnj}\psi _{omnp}\psi
^{oklp}\,.  \label{4-th}
\end{eqnarray}%
The ratios $I_{12}^{(4)}/(I^{(2)})^{2}$, $I_{13}^{(4)}/(I^{(2)})^{2}$,
and $I_{14}^{(4)}/(I^{(2)})^{2}$ are, in addition, invariant with
respect to multiplication of the state vector by an arbitrary complex
constant. Were these ratio linearly independent, they would suffice
for a complete characterization of four-qubit entanglement.  However,
they are not. The following identity,
\begin{equation}
I_{12}^{(4)}+I_{13}^{(4)}+I_{14}^{(4)}\ =\ \frac{3}{2}\left(
I^{(2)}\right)^{2}\,, \label{identJ}
\end{equation}%
makes such quantities inconvenient for entanglement characterization.

Thus, it is necessary to turn to the sixth-order invariants.  We
consider the following three independent combinations,
\begin{eqnarray}
\hspace*{-2cm}I_{12}^{(6)}\ & =\frac{1}{6}\ \left( \psi _{ingd}\psi
_{mrko}\psi _{sjph}-\psi _{ingo}\psi _{mrkh}\psi _{sjpd}\right) \psi
^{mrgd}\psi ^{inph}\psi ^{sjko}\,, \nonumber \\
\hspace*{-2cm}I_{23}^{(6)}\ & =\frac{1}{6}\ \left( \psi _{ijpo}\psi
_{mngh}\psi _{srkd}-\psi _{ijpd}\psi _{mngo}\psi _{srkh}\right) \psi
^{mrgd}\psi ^{inph}\psi ^{sjko}\,, \nonumber \\
\hspace*{-2cm}I_{13}^{(6)}\ & =\frac{1}{6}\ \left( \psi _{ijkh}\psi
_{mnpd}\psi _{srgo}-\psi _{ijgh}\psi _{mnkd}\psi _{srpo}\right) \psi
^{mrgd}\psi ^{inph}\psi ^{sjko}\,, \label{6-th}
\end{eqnarray}
whose differences give the invariants of Eq.~(\ref{4-th}) multiplied
by $I^{(2)}$. The explicit form of these invariants for a generic
state is awkward. However, they take a simple form for the canonic
state under $sl$-transformations, which allows us to explicitly relate
them to the canonic amplitudes.  One finds
\begin{eqnarray}
\hspace*{-2cm}\psi _{0000}&
=\frac{\sqrt{\sqrt{I_{13}^{(6)}+Q}+\sqrt{I_{23}^{(6)}+Q}+\sqrt{%
I_{12}^{(6)}+Q}-\left( I^{(2)}\right) ^{3/2}}}{\sqrt{2}\left(
I^{(2)}\right) ^{1/4}}, \nonumber \\
\hspace*{-2cm}\psi _{1100}& =\psi
_{0011}=\frac{\sqrt{\sqrt{I_{13}^{(6)}+Q}-\sqrt{%
I_{23}^{(6)}+Q}-\sqrt{I_{12}^{(6)}+Q}+\left( I^{(2)}\right)
^{3/2}}}{2\left( I^{(2)}\right) ^{1/4}}, \nonumber \\
\hspace*{-2cm}\psi _{1001}& =\psi
_{0110}=\frac{\sqrt{\sqrt{I_{23}^{(6)}+Q}-\sqrt{%
I_{13}^{(6)}+Q}-\sqrt{I_{12}^{(6)}+Q}+\left( I^{(2)}\right)
^{3/2}}}{2\left( I^{(2)}\right) ^{1/4}}, \nonumber \\
\hspace*{-2cm}\psi _{0101}& =\psi
_{1010}=\frac{\sqrt{\sqrt{I_{12}^{(6)}+Q}-\sqrt{%
I_{23}^{(6)}+Q}-\sqrt{I_{13}^{(6)}+Q}+\left( I^{(2)}\right)
^{3/2}}}{2\left( I^{(2)}\right) ^{1/4}}, \nonumber \\
\hspace*{-2cm}\psi _{1111}&
=\frac{\sqrt{I_{13}^{(6)}+Q}+\sqrt{I_{23}^{(6)}+Q}+\sqrt{%
I_{12}^{(6)}+Q}+\left( I^{(2)}\right) ^{3/2}}{2\sqrt{2}\left(
I^{(2)}\right)
^{1/4}\sqrt{\sqrt{I_{13}^{(6)}+Q}+\sqrt{I_{23}^{(6)}+Q}+\sqrt{I_{12}^{(6)}+Q}%
-\left( I^{(2)}\right) ^{3/2}}}, \label{AMPL}
\end{eqnarray}%
where $Q$ is a root of the following cubic equation: 
\begin{equation}
(I_{13}^{(6)}+Q)(I_{23}^{(6)}+Q)(I_{12}^{(6)}+Q)=( I^{(2)} )^{3}Q^{2}\,.  
\label{QE}
\end{equation}%
The set of Eqs.~(\ref{AMPL}) determines the canonic state vector form
with respect to pure $SL$-transformations.  Upon dividing
Eqs.~(\ref{AMPL}) by $\psi _{0000},$ the ratios $\psi _{1100}/\psi
_{0000}$, $\psi _{1001}/\psi _{0000}$, and $\psi _{0101}/\psi _{0000}$
respectively yield the $sl$-tanglemeter coefficients $\beta _{3}$,
$\beta _{5}$, and $\beta _{6}$, which are also
scaling-invariant. Different roots of the cubic equation (\ref{QE})
yield different $sl$-canonic states related by $SL$
transformations. We can choose one particular root by minimizing the
difference between the normalization of the canonic state and the
initial normalization. Thus, as conjectured in Sec.~\ref{secII}, the
$sl$-entanglement in the four-qubit assembly may be completely
characterized by {\em three independent scale-invariant complex
ratios},
\begin{eqnarray}
\hspace*{-1cm}\beta _{3}&
=\frac{\sqrt{\sqrt{I_{13}^{(6)}+Q}-\sqrt{I_{23}^{(6)}+Q}-\sqrt{%
I_{12}^{(6)}+Q}+\left( I^{(2)}\right) ^{3/2}}}{\sqrt{2}\sqrt{\sqrt{%
I_{13}^{(6)}+Q}+\sqrt{I_{23}^{(6)}+P}+\sqrt{I_{12}^{(6)}+Q}-\left(
I^{(2)}\right) ^{3/2}}}, \nonumber \\ 
\hspace*{-1cm} \beta _{5}&
=\frac{\sqrt{\sqrt{I_{23}^{(6)}+Q}-\sqrt{I_{13}^{(6)}+Q}-\sqrt{%
I_{12}^{(6)}+Q}+\left( I^{(2)}\right) ^{3/2}}}{\sqrt{2}\sqrt{\sqrt{%
I_{13}^{(6)}+Q}+\sqrt{I_{23}^{(6)}+Q}+\sqrt{I_{12}^{(6)}+Q}-\left(
I^{(2)}\right) ^{3/2}}}, \nonumber \\ 
\hspace*{-1cm}\beta _{6}&
=\frac{\sqrt{\sqrt{I_{12}^{(6)}+Q}-\sqrt{I_{23}^{(6)}+Q}-\sqrt{%
I_{13}^{(6)}+Q}+\left( I^{(2)}\right) ^{3/2}}}{\sqrt{2}\sqrt{\sqrt{%
I_{13}^{(6)}+Q}+\sqrt{I_{23}^{(6)}+Q}+\sqrt{I_{12}^{(6)}+Q}-\left(
I^{(2)}\right) ^{3/2}}}, \label{betas}
\end{eqnarray}%
emerging from the invariants of Eqs.~(\ref{6-th})-(\ref{J2}). In view
of this, a natural measure of $sl$-entanglement is provided by the sum
of squared moduli of the $sl$-tanglemeter coefficients $\beta$,
$\mathcal{S}_{2}=\sum \left\vert \beta \right\vert ^{2}$. This yields
$\mathcal{S}_{2}=$ $0$ for the GHZ canonic state, whereas
$\mathcal{S}_{2}\not=$ $0$ for all other states, thereby exhibiting a
similar behavior to the hyper-determinant \cite{Akimasa}. Accordingly,
this measure quantifies how close the orbit is to the {GHZ}-orbit. The
quantity $\sum \left\vert \beta -\beta ^{\prime }\right\vert ^{2}$ may
likewise serve as a measure characterizing the distance between two
different $sl$-orbits.

As a next question, we wish to suggest a simple measure for
characterizing $su$-entanglement in four qubits. A natural candidate
is the sum $\mathcal{S}_{1}=\sum \left\vert \psi \right\vert ^{2}$
over the probabilities in Eq.~(\ref{AMPL}), which gives the standard
normalization of the canonic-like state. Once the invariants of
Eqs.~(\ref{J2})-(\ref{6-th}) are calculated for a state with unit
normalization, this sum quantifies the extent by which the $SL$
transformation required for setting the state to the canonic form
differs from a unitary transformation.  Thus, $\left\vert \ln \sum
\left\vert \psi \right\vert ^{2}\right\vert$ provides us with a
suitable measure of such a non-unitarity. By construction, the latter
quantity is able to discriminate between different $su$-orbits that
belong to the same $sl$-orbit.  Interestingly, as found in
\cite{nilpotent}, this measure exhibits a strong correlation with the
quadratic $sl$-invariant $I^{(2)}$.

\subsection{Measures for fourpartite entanglement}

Having suggested $sl$- and $su$-measures for four qubits in terms of
the tanglemeter's coefficients, we finally proceed to address the more
delicate issue of constructing a measure of genuine fourpartite
entanglement \cite{Emary}. In addition to behaving as an entanglement
monotone, such a measure should satisfy the requirement of being zero
in the $sl$-orbits that do not bear genuine fourpartite
entanglement. Within constructions based on $sl$-invariants (different
approaches have also been suggested, see e.g. \cite{ Emary,
Osterloh,Levay1}), the combination of invariants able to satisfy the
last requirement is not known to date.  For example, $I^{(2)}$ in
Eq.~(\ref{J2}) is a low-order entanglement monotone, but it cannot
serve as a good fourpartite measure since it attains its maximum value
$1$ for both the four-qubit GHZ state and for a product of two Bell
pairs, that is a four-qubit state which manifestly contains no genuine
fourpartite correlations \cite{Verstraete2}.  The so-called
$4$-concurrence introduced in \cite{Wong}, that is just
$\left|I^{(2)}\right|$, exhibits a similar unfavorable behavior.  On
the other hand, the hyper-determinant $\Delta $ \cite{Akimasa} is
nonzero in the general family of orbits $G_{abcd}$, and zero in all
others as well as in the GHZ orbit. According to our results (Table
I), recall that the families of orbits $L_{abc_{2}} $,
$L_{a_{2}b_{2}}$, and $L_{a_{2}0_{3\oplus 1}}$ are derived as special
cases of the general family, and also contain genuine fourpartite
entanglement in general.

Observing that the determinant ${\cal D}_4$ of the infinitesimal
transformations given in Eq.~(\ref {EQ62e}) is precisely equal to zero
in the orbits $G_{abcd}$, $L_{abc_{2}}$, $L_{a_{2}b_{2}}$, and
$L_{a_{2}0_{3\oplus 1}}$, we express it in terms of the canonic state
amplitudes,
\begin{equation}
\kappa_4 =\left\vert 
\begin{array}{cccc}
\Psi _{15} & 2\psi _{6}\psi _{10} & 2\psi _{6}\psi _{12} & 2\psi _{10}\psi
_{12} \\ 
2\psi _{5}\psi _{9} & \Psi _{15} & 2\psi _{5}\psi _{12} & 2\psi _{9}\psi
_{12} \\ 
2\psi _{3}\psi _{9} & 2\psi _{3}\psi _{10} & \Psi _{15} & 2\psi _{9}\psi
_{10} \\ 
2\psi _{3}\psi _{5} & 2\psi _{3}\psi _{6} & 2\psi _{5}\psi _{6} & \Psi _{15}%
\end{array}%
\right\vert/\psi_0^8 ,
\end{equation}%
where $\Psi _{15}=$ $-\psi _{15}\psi _{0}+\psi _{6}\psi _{9}+\psi
_{3}\psi _{12}+\psi _{5}\psi _{10}$.  Our proposal is to consider the
quantity
\begin{equation}
\mathcal{K}_4 =4 \sqrt{\left\vert
\kappa_4 \right\vert/\mathcal{A}^4}\, 
\end{equation}
as a measure of proper fourpartite entanglement.  Here, the
normalization $\mathcal{A}=\sum_i\left\vert\psi _{i}/\psi
_{0}\right\vert^2$ ensures that $\mathcal{K}_4$ consistently ranges
between $0,1$.  Note that $\mathcal{K}_4$ is constructed as a function
of $su$-canonic amplitudes, thus it remains invariant under local
unitary transformations, while it is not a $sl$-invariant.

\begin{figure}[h]
\begin{center}
{\centering{\includegraphics*[width=0.55\textwidth]{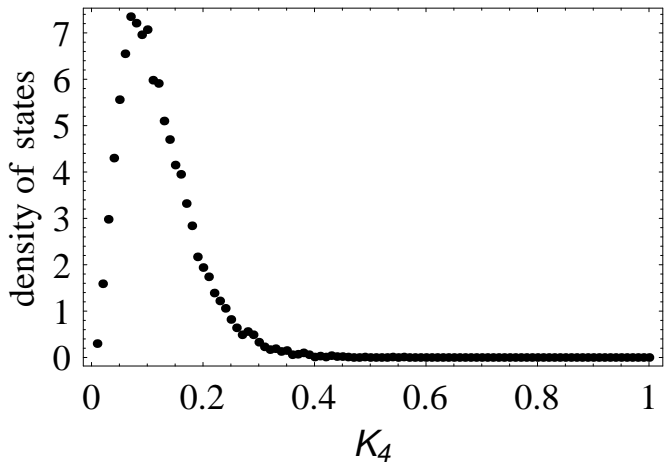}}} 
\end{center}
\caption{The distribution of $\mathcal{K}_4$ for a set of $10^4$
four-qubit pure states sampled uniformly over the Haar measure. 
See for comparison Fig. 2 (inset)
in \cite{Kendon}. }
\label{Fig0}
\end{figure}

In order to gain further insight into the properties of
$\mathcal{K}_4$, its behavior on a typical four-qubit state is
numerically investigated.  The distribution of $\mathcal{K}_4$ for
pure states sampled uniformly over the Haar measure is depicted in
Fig.~\ref{Fig0}.  A non-monotonic behavior is clearly seen, 
peaked around the value $0.1$, qualitatively resembling the
distribution of the $3$-tangle for three-qubit pure states
\cite{Kendon}.  Unlike the $4$-concurrence, $\mathcal{K}_4$ attains
its maximum value $1$ for the GHZ state, and gives zero for all the
states which are separable in some way.  For a four qubit cluster
state \cite{Briegel}, for instance, it gives $\mathcal{K}_4= 0.3265$,
which is well above the average value.
 
\vspace*{3mm}

While the above features make $\mathcal{K}_4$ an attractive candidate
for quantifying fourpartite entanglement, a main disadvantage of
$\mathcal{K}_4$ is that it inherits the redundancy of our
classification, vanishing whenever the general class of orbits cannot
be reached by infinitesimal transformations -- irrespective of whether
it might be reached by finite transformations.  Furthermore, the
calculation of $\mathcal{K}_4$ for a given pure state requires in
general that the latter is first reduced to its $su$-canonic form. On
the other hand, extending the construction of this measure to $n>4$
qubits is relatively straightforward in principle.  For $n=4$, the
fact that the determinant $\kappa_4$ does not contain the
second-highest order terms is a sign that this measure is approximate
for arbitrary states. However, this effect may expected to become less
pronounced (hence the accuracy of such approximation improves) with
increasing $n$.  Finally, it remains an open question to ascertain
whether this measure is an entanglement monotone.

\section{Discussion}
\label{secIV}

In summary, we have demonstrated how the approach based on nilpotent
polynomials may be employed to identify entanglement classes for the
illustrative yet highly nontrivial situation of four qubits in a pure
state. Even if the approach is somewhat redundant compared to
mathematically more sophisticated methods, we believe it has the
advantage of offering a clearer physical interpretation, and it may
also be extended straightforwardly to larger multipartite ensembles
and higher-dimensional subsystems.

In the context of the resulting classification, we have suggested
additional class of measures beside the existing ones, which remain
invariant under either local unitary ($su$) or arbitrary local
invertible ($sl$) transformations.  We employ the nilpotent invariant
coefficients for the construction of such measures as an alternative
route to invoking polynomial algebraic invariants.

Finally, we have also suggested a measure of genuine fourpartite
entanglement.  Our prospective measure is both, by construction, a
$su$-invariant and it vanishes on the special orbits where no genuine
fourpartite entanglement exists.  We also note that while it would be
very illustrative to apply the nilpotent method in the well-explored
three qubit case, the procedure implemented here is not viable, since
in this case the $SL$ coset dimension $D_{sl}<0$ and, consequently,
the dynamical set of equations employed in the analysis becomes
degenerate.

It is our hope that the results presented here may serve as a stimulus
to prompt further investigations and applications of the nilpotent
polynomial formalism as a tool exploring entanglement.

\section*{Acknowledgments}

It is a pleasure to thank Vladimir Akulin and Andrei Smilga for
insightful discussions and collaboration on the nilpotent entanglement
program.  The authors are also indebted to Winton G. Brown for
valuable suggestions and a critical reading of the manuscript.  AM
gratefully acknowledges Ile de France for financial support.

\end{document}